\documentclass[12pt]{article}
\usepackage{graphicx}
\usepackage{amssymb}
\usepackage{amsfonts}

\title{Role of qubits in quantum entanglement and quantum teleportation}
\author{ Laure Gouba \\
{\em
The Abdus Salam International Centre for Theoretical Physics (ICTP)}\\
{\em Strada Costiera 11, I-34151 Trieste Italy.}\footnote {\em Contact: laure.gouba@gmail.com}}

\begin{document}

\maketitle

\begin{abstract}
\noindent
 A qubit is an exhibition of quantum entanglement and a key element in the quantum teleportation process. In this paper, we review the role of qubits in quantum
entanglement and quantum teleportation
\end{abstract}


\section{Introduction}\label{sec1}

Back to almost ninety years ago, Erwin Schr\"odinger coined the name {\it verschr\"anking} to a correlation of quantum nature \cite{schro, schr}. The word {\it verschr\"anking}, casually used in German for non-physicists only in the sense of {\it folding the arms}, has been loosely translated to {\it entanglement} with more inspiring connotations and its meaning has changed its flavour over the decades \cite{susskind}.

The question of expected locality of the entangled quantum systems raised by EPR \cite{epr1,epr2} allowed John Stewart Bell to discover his famous inequalities, serving as a test and demonstration of strange properties of the simplest entangled wave function represented by a singlet state \cite{Bell1, Bell2, Bell3}.  Still, one had to wait long for the proposals and practical applications of quantum entanglement. 

Until $1975$, a decisive experiment based on the violation of Bell's inequalities and verifying the veracity of quantum entanglement was missing. The experiment led by French physicist Alain Aspect at the \'Ecole Sup\'erieure d'Optique in Orsay between $1980$ and $1982$ was the first quantum mechanics experiment to demonstrate the violation of Bell's inequalities \cite{aspect1, aspect2}. This experiment is called the Aspect's experiment. It confirmed the predictions of quantum mechanics and thus confirmed its incompatibilities with local theories. 

Quantum entanglement is a phenomenon that has no counterpart in classical physics. It can be seen as the most non-classical feature of quantum mechanics that has raised numerous philosophical, physical and mathematical questions since the early days of the quantum theory \cite{niel}. 
Entanglement is a quantum mechanical form of correlation which appears in many areas, such as condensed matter physics, quantum chemistry and other areas of physics \cite{fazio}. 
Applications of quantum entanglement include quantum teleportation, superdense coding, quantum information theory, quantum cryptography and quantum computation \cite{niel, 4horodecki, pirandola}.
In this paper, we feature quantum teleportation.
The term teleportation comes from science fiction, meaning to make a person or object disappear while a replica appears somewhere else. Quantum teleportation was suggested by Bennett et al. \cite{bennett}, where the process, unlike some science fiction versions, defies no physical laws. In particular, it cannot take place instantaneously or over a spacelike interval because it requires, among other things, sending a classical message from a sender to a receiver. Quantum teleportation is recognized nowadays as an important application of quantum entanglement in correlation with quantum information processing.

As the simplest system to display quantum entanglement is the system of two qubits, and quantum teleportation involves transmitting the quantum state of a qubit from one location to another using entanglement and classical communication, 
in this paper, we review the role of qubits in quantum entanglement and quantum teleportation. Section \ref{sec2} is about qubits, and then we present in section \ref{sec3} the criteria of separability of entanglement in bipartite systems. The protocol of quantum teleportation and the fidelity of teleportation are exhibited in section \ref{sec4}. Concluding remarks are given in \ref{sec5}.

\section{Qubits}\label{sec2}

The quantum analogue of a system with n states is the n-dimensional Hilbert space $\mathbb{C}^n$. A quantum system with $n =2$ states is called a qubit (quantum bit). Just as a classical bit can have a state of either $0$ or $1$, the two most common states for a qubit (quantum bit) are the states 
\begin{equation}
|0\rangle = \left(\begin{array}{c}
1\\0
\end{array}\right)
\quad \textrm{and}\quad |1 \rangle = \left(\begin{array}{c}
0\\1
\end{array}\right).
\end{equation}  
Any two-level quantum-mechanical system can be used as a qubit. The qubits describe the simplest two-state quantum mechanical system and are allowed to be in a coherent superposition of both states simultaneously. This means that the state vectors in $\mathcal{H} = \mathbb{C}^2$ are of the form 
\begin{eqnarray}\nonumber
|v\rangle & = & \alpha \vert 0\rangle + \beta \vert 1 \rangle,\\\nonumber
{}\\
& = & 
\left(\begin{array}{c}
\alpha\\ 0
\end{array} \right) + \left(\begin{array}{c}
0\\ \beta
\end{array} \right)  = \left(\begin{array}{c}
\alpha\\ \beta
\end{array} \right)
\end{eqnarray} 
for some $\alpha, \beta \in \mathbb{C}$, with $|\alpha|^2 + |\beta |^2 = 1$. The coefficients $\alpha$ and $\beta$ can be viewed as real number without loosing. 

\begin{itemize}
\item Quantum superposition principle:
{\it If a quantum system can be in the state $\vert 0\rangle$ and can also be in the state $|1\rangle$, then quantum mechanics allows the system to be in any arbitrary state $|v\rangle = \alpha |0\rangle + \beta |1\rangle $. The state $|v\rangle$ is said to be in a superposition of $|0\rangle$ and $|1\rangle$ with probability amplitude $\alpha$ and $\beta$.}
Two famous states are given by 
\begin{equation}
|+\rangle = \frac{1}{\sqrt 2}|0\rangle + \frac{1}{\sqrt 2}|1\rangle, \quad 
|-\rangle = \frac{1}{\sqrt 2}|0\rangle - \frac{1}{\sqrt 2} |1 \rangle.
\end{equation}
\item How the qubits are realized physically?
\begin{enumerate}
\item The qubits might be represented by two states of an electron orbiting an atom.
\item The qubits might be  represented by two directions of the spin of a particle: for example to measure the spin of a particle along z-axis, it is Up or Down, that is ($z+$ direction) and ($z-$ direction) or $|\uparrow\rangle $ and $|\downarrow\rangle$. For computational purpose it is convenient to use $|0\rangle $ and $|1\rangle$. 
\item The qubits might be represented by two polarizations of a photon.
\item  The qubits might be realized in superconducting transmon qubits \cite{qubitRe1, qubitRe2}.
\end{enumerate}
\item The computational basis $\{|0\rangle,\; |1\rangle\}$ is typically used to represent two exclusive states of a quantum system used as quantum-$0$ and quantum-$1$. For example, if the energy of an electron in an atom is used as our quantum bit, we could say that the ground state (lowest energy) is our quantum-$0$ and an excited state (higher energy) is our quantum-1.  Since the ground state and the excited states are mutually exclusive, the representation could be: ground state $\leftrightarrow$ $|0\rangle = \left[\begin{array}{c}
1\\0
\end{array}\right]$, excited state  $\leftrightarrow$ $|1\rangle = \left[\begin{array}{c}
0\\1
\end{array}
\right]$.
\end{itemize}
Quantum mechanics allows the qubit to be in a coherent superposition of multiple states simultaneously, a property that is fundamental to quantum computing. The qubit is an exhibition of quantum entanglement in the sense that quantum entanglement is a property of two or more qubits that allows a set of qubits to express higher correlation than is possible in classical systems.  Benjamin Schumacher coined the term qubit, that has been created in jest during a conversation with William Wootters \cite{benschu}.

\section{Quantum entanglement}\label{sec3}

The fundamental question in quantum entanglement theory is which states are entangled and which are not. This question, known as the separability problem, has been solved for pure states \cite{neven} and for $2 \times 2$ and $2 \times 3$ systems \cite{horo1}, but it remains an open problem from both a theoretical and experimental standpoint.

A separability condition can be necessary or necessary and sufficient conditions for separability \cite{dag}. A necessary condition for separability has to be fulfilled by every separable state. In that case, if a state does not fulfill the condition, it has to be entangled, but if it fulfills we cannot conclude.

The wave function describing a quantum system is entangled if it cannot be written as a tensorial product of states of subsystems. The simplest example is the singlet state of two spin-$\frac{1}{2}$ particles \cite{bohm}
\begin{equation}
|\psi\rangle = \frac{1}{\sqrt{2}}\left( |0\rangle \otimes|1\rangle - |1\rangle \otimes |0\rangle \right).
\end{equation}
We can define $|01\rangle = |0 \rangle \otimes |1\rangle$ and 
$|10\rangle = |1 \rangle \otimes |0\rangle$.
It can be proven that $|\psi\rangle \neq |\varphi\rangle \otimes |\phi\rangle$ for any $|\varphi\rangle$ and $|\phi\rangle$ describing subsystems, $|0\rangle$ standing for ``spin up" state and $|1\rangle$ standing for ``spin down" state. 
The `` nonfactorisability" of any bipartite pure state implies that its reduced density matrices are mixed. The above definition is naturally generalized to the entanglement of multiparticle pure state. A bipartite pure quantum state $|\psi\rangle_{AB}\in \mathcal{H}_A \otimes \mathcal{H}_B$ is called entangled when it cannot be written as 
\begin{equation}
|\psi\rangle_{AB} = | \psi\rangle_A \otimes |\psi\rangle_B,
\end{equation} 
 for some $|\psi\rangle_A \in \mathcal{H}_A$ and $|\psi\rangle_B \in \mathcal{H}_B$. A mixed state or density matrix $\rho_{AB}$ which is semi-definite operator on $\mathcal{H}_A \otimes \mathcal{H}_B$ is called entangled when it cannot be written in the following form 
\begin{equation}
 \rho = \sum_i p_i|\psi_i\rangle_A{}_A\langle \psi_i| \otimes |\psi_i\rangle_B{}_B\langle \psi_i|
\end{equation}
 Here the coefficients $p_i$ are probabilities, that means $0\le p_i \le 1$ and $\sum_{i} P_i = 1 $. Note that in general neither $\{|\psi_i\rangle_A\}$
 nor $\{|\psi_i\rangle_B\}$ have to be orthogonal.

\subsection{Bell-CHSH inequalities}

The Bell inequality was originally designed to test predictions of quantum mechanics against those of a local hidden variables theory \cite{Bell1}. 
Bell's inequalities were initially dealing with two qubits, i.e, two-level systems, and provide a necessary criterion for the separability of 2-qubits states. For pure states, Bell's inequalities are also sufficient for separability.  It has been proven by Gisin that any non-product state of two-particle systems violates a Bell inequality \cite{gisin}. 
This inequality, which involves three vectors in real space $\mathbb{R}^3$ determining which component of a spin to be measured by each party, has been extended for the case involving four vectors by Clauser, Horne, Shimony and Holt (CHSH) in 1969 \cite{chshin}. The Bell-CHSH inequality also provides a test to distinguish entangled from non-entangled states.

Consider a system of two qubits. Let $A$ and $A'$ denote observables on the first qubit, $B$ and $B'$ denote observables on the second qubit. The Bell-CHSH inequality says that for non-entangled states, meaning for states of the form $\rho = \rho_1\otimes \rho_2$, or mixtures of such states, the following inequality holds:
\begin{equation}\label{bin1}
| \langle\; A\otimes B + A\otimes B' + A'\otimes B - A'\otimes B'\;\rangle_\rho|\le 2,
\end{equation}
where $\langle A\otimes B \rangle_{\rho} := \textrm{Tr}\; \rho (A\otimes B)$ and 
$\langle A\otimes B \rangle_{\psi} = \langle \psi |A\otimes B|\psi\rangle $ for the expectation value of $A \otimes B$ in the mixed states $\rho$ or pure state $|\psi\rangle$.
As an example, we consider a two qubits state $|\phi\rangle = \frac{1}{\sqrt{2}} (|00\rangle + |11\rangle)$, and the observables 
\begin{equation}
A = \frac{1}{\sqrt{2}}(\sigma_x +\sigma_z),\: A' = \frac{1}{\sqrt{2}}(\sigma_x - \sigma_z), \: B = \sigma_x, \: B' = \sigma_z,
\end{equation}
where $\sigma_x$ and $\sigma_z$  are Pauli matrices. We have then explicitly
\begin{equation}
\sigma_x = \left(\begin{array}{cc}
0 & 1\\
1 & 0
\end{array} \right); \quad \sigma_z = \left(\begin{array}{cc}
1  & 0\\
0 & -1
\end{array} \right); 
\end{equation}
and
\begin{equation}
A  = \frac{1}{\sqrt 2}\left(\begin{array}{cc}
1 & 1\\
1 & -1
\end{array} \right);\: A' = \frac{1}{\sqrt 2}\left(\begin{array}{cc}
-1 & 1\\
1 & 1
\end{array} \right);\: B = \left(\begin{array}{cc}
0 & 1\\
1 & 0
\end{array} \right); \: B' = \left(\begin{array}{cc}
1 & 0\\
0 & -1
\end{array} \right).
\end{equation}
It is easy to check that 
\begin{equation}\label{eqappend}
\langle \phi|\; A\otimes B + A\otimes B' + A'\otimes B - A'\otimes B'\;|\phi\rangle = 2\sqrt 2.
\end{equation}
The demonstration of the equation (\ref{eqappend}) is given in the Appendix.
The state $|\phi\rangle$, which violates the Bell-CHSH inequality, is a well-known entangled state, and is one of the Bell pairs, a maximally entangled state.
The maximal violation of (\ref{bin1}), for entangled states, follows from an inequality of Cirelson \cite{cirelson}
\begin{equation}\label{bin2}
| \langle A\otimes B + A\otimes B' + A'\otimes B - A'\otimes B'\rangle_\rho|\le 2\sqrt 2.
\end{equation}
The equality in equation (\ref{bin2}) can be attained by the singlet state.
Historically, Bell-CHSH inequalities were the first tool for the recognition of entanglement; however, it is well-known for some time that the violation of a Bell-CHSH inequality is only a sufficient condition for entanglement and not a necessary one, and that there are many entangled states that satisfy them \cite{werner}.
Bell-CHSH inequalities were generalized to N qubits, whose violations provide a criterion to distinguish the separable states from the entangled states \cite{gbin1,gbin2, gbin3}.

\subsection{Criteria of separability}

An overview of the progress in the separability problem in two qubits system has been investigated in \cite{gnonfin} where the following criteria have been discussed.
\begin{enumerate}
\item  Schmidt decomposition criterion \cite{sch, eker, peres, peres1}.
\item  The Positive Partial Transpose (PPT) criterion 
\cite{ horo1, peres1, peres2, phoro}.
\item  Entropy of entanglement criterion.
\item The reduction criterion \cite{horo1, cerf}.
\item Concurrence criterion \cite{woo1, woo2, woo3,zhao}.
\item  The majoration criterion \cite{marshall, nielsen}.
\item The computable cross norm or realignment (CCNR) criterion \cite{rudolph1, rudolph2, chenwu}.
\item A matrix realignment criterion \cite{chenwu,vicente}.
\item The correlation matrix criterion \cite{vicente, block, hioe, fano}.
\item  Enhanced entanglement criterion via SIC POVMs \cite{shang, icpovm}.
 \item  Positive maps criterion \cite{pm1, pm2, pm3, pm4, pm5}.
\item  The entanglement witnesses \cite{ horo1, terhal1, guhne, ganguly, dagmar}.
\item  Local uncertainty relations (LURs) criterion \cite{lur1,lur2,lur4, lur5}.
\item  The Li-Qiao criterion \cite{horo1,vicente,liqiao,horn1, horn2, horn3}.
\end{enumerate}

\section{Quantum teleportation}\label{sec4}

Quantum teleportation involves transferring complete information of one quantum state from one location to another location with the aid of long-range EPR correlations in an entangled state.  This entangled state is shared between the two parties, which are known as the sender and the receiver. At first, the sender makes some measurements with the information state and her or his shared part of the entangled state. In this process, the information state disappears at the sender’s end and instantly appears at the receiver’s end. This is obtained when the receiver makes some unitary transformation that depends on some result of the sender’s measurement, which is received through some classical channel. The transmission of quantum states can be accomplished without using any entanglement \cite{noentang1, noentang2}, whether it is only classical communication or by the transmission of classical bits, where the sender and the receiver share entanglement. We are interested in the second situation. In that sense, quantum teleportation involves transmitting the quantum state of a qubit from one location to another using entanglement and classical communication. It's not about physically moving the qubit, but rather transferring its quantum information. An illustration is given in the study of the effect of nonmaximality of a quasi-Bell state-based quantum channel in the context of the teleportation of a qubit \cite{aremua-gouba, lgouba}.

\subsection{Teleportation through a nonmaximally entangled channel}

The teleportation protocol begins with a qubit $ |\psi \rangle_C $, in Amy's possession, that she wants to send to Bouba. This  is 
\begin{equation}\label{sc}
|\psi\rangle_C = \cos\frac{\theta}{2}|0 \rangle 
+ \sin\frac{\theta}{2}\exp{(i\phi)}|1 \rangle.
\end{equation} In the protocol, Amy and Bouba share this nonmaximal entangled channel
\begin{equation}\label{sab}
|\psi\rangle_{AB} = \alpha |00\rangle + \beta |11\rangle,
\end{equation}
where we assume that $\alpha$ and $\beta$ are real and that $ \alpha \le \beta$.  The subscripts  A and B in the entangled state refer to Amy's or Bouba's state, while the subscript C refers to the initial state in Amy's possession. Since the entangled channel shared by Amy and Bouba is nonmaximal, it is not possible to perform this qubit teleportation perfectly, so we would like to compute the fidelity. 
Amy obtains one of the states in the pair, with the other going to Bouba. At this point, Amy has two states: the one she wants to teleport, and A, one of the entangled pairs, and Bouba has one particle, B.  The state of the total system, consisting of the input qubit in the equation (\ref{sc}) and the shared entanglement in the equation (\ref{sab}) becomes  
\begin{eqnarray}\nonumber
|\psi\rangle_C \otimes |\psi\rangle_{AB} &= & (\cos\frac{\theta}{2}|0 \rangle_C
+ \sin\frac{\theta}{2}\exp{(i\phi)}|1 \rangle_C)
\\
&\otimes& ( \alpha |0\rangle_A|0\rangle_B + \beta |1\rangle_A |1\rangle_B).
\end{eqnarray}
The initial state is therefore
\begin{eqnarray}\nonumber
|\psi \rangle_{CAB}  &= & \alpha\cos\frac{\theta}{2}\, |0\rangle_C|0\rangle_A |0 \rangle_B
+ \beta \cos\frac{\theta}{2}\, |0\rangle_C
|1\rangle_A |1 \rangle_B\\
&+& \alpha\sin\frac{\theta}{2}\, e^{i\phi}\,
|1\rangle_C|0\rangle_A |0 \rangle_B
+ \beta \sin\frac{\theta}{2}\, e^{i\phi}\,
| 1 \rangle_C
|1\rangle_A |1 \rangle_B.
\end{eqnarray}
Amy makes a local measurement on the Bell basis  on the two states in her possession. In order to make the result of her measurement clear, it is best to write  the state of Amy's two qubits as superposition of the Bell basis as follows
\begin{eqnarray}
|0\rangle_C|0\rangle_A & =& \frac{1}{\sqrt{2}}
(|\phi^+\rangle_{CA} + |\phi^-\rangle_{CA}); \\
|0\rangle_C |1\rangle_A & =& \frac{1}{\sqrt{2}}
(|\psi^+\rangle _{CA} + |\psi^-\rangle_{CA}); 
\\
|1 \rangle_C| 0 \rangle_A & =& \frac{1}{\sqrt{2}}
(|\psi^+\rangle_{CA}  - |\psi^-\rangle_{CA});
 \\
| 1\rangle_C | 1 \rangle_A & =& \frac{1}{\sqrt{2}}
(|\phi^+ \rangle_{CA} - |\phi^-\rangle_{CA}).
\end{eqnarray}
The state of the total system A, B and C together form the following four-term superposition.
\begin{eqnarray}\nonumber
|\psi \rangle_{CAB} &=& \frac{1}{\sqrt{2}}
\left[
|\phi^+ \rangle_{AC}\otimes \left( 
\alpha\cos\frac{\theta}{2}\,|0\rangle_B + \beta\sin\frac{\theta}{2}e^{i\phi}|1\rangle_B\right)\right.\\\nonumber
&+& \left.
|\phi^- \rangle_{AC}\otimes \left( 
\alpha\cos\frac{\theta}{2}\,|0\rangle_B - \beta\sin\frac{\theta}{2}e^{i\phi}|1\rangle_B\right)\right.\\\nonumber
&+& \left.
|\psi^+ \rangle_{AC}\otimes \left( 
\beta\cos\frac{\theta}{2}\,|1\rangle_B + \alpha \sin\frac{\theta}{2}e^{i\phi}|0\rangle_B\right)\right.\\
&+&\left.
|\psi^- \rangle_{AC}\otimes \left( 
\beta\cos\frac{\theta}{2}\,|1\rangle_B - \alpha\sin\frac{\theta}{2}e^{i\phi}|0\rangle_B\right)
\right].
\end{eqnarray}
The three states which are the two states in Amy's possesion and one of Bouba's, are still in the same total state.
All the three states are still in the same total state, since no operations have been performed. The above equation is just a change of basis on Amy's part of the system. This change has moved the entanglement from particles A and B to particles C and A.
The teleportation occurs when Amy measures her two qubits in the Bell basis:
$|\phi^+\rangle_{CA},\, 
|\phi^-\rangle_{CA},\, |\psi^+\rangle_{CA},\, 
|\psi^-\rangle_{CA}$.
The result of Amy's local measurement is a collection of two classical bits $00,\: 01,\: 10,\: 11$ related to one of the four states after the three-particle state has collapsed into one of the states:
\begin{eqnarray}
&{}& |\phi^+ \rangle_{AC}\otimes \left( 
\alpha\cos\frac{\theta}{2}\,|0\rangle_B + \beta\sin\frac{\theta}{2}e^{i\phi}|1\rangle_B\right);\\
&{}&
|\phi^- \rangle_{AC}\otimes \left( 
\alpha\cos\frac{\theta}{2}\,|0\rangle_B - \beta\sin\frac{\theta}{2}e^{i\phi}|1\rangle_B\right);\\
&{}&
|\psi^+ \rangle_{AC}\otimes \left( 
\beta\cos\frac{\theta}{2}\,|1\rangle_B + \alpha \sin\frac{\theta}{2}e^{i\phi}|0\rangle_B\right);\\
&{}&
|\psi^- \rangle_{AC}\otimes \left( 
\beta\cos\frac{\theta}{2}\,|1\rangle_B - \alpha\sin\frac{\theta}{2}e^{i\phi}|0\rangle_B\right).
\end{eqnarray}
Amy sent her result to Bob through a classical channel.  Two classical bits can communicate which of the four results she obtained. Her two states are now entangled to each other in one of the four Bell states, and the entanglement originally shared between Amy's and Bouba's state is now broken. 
Bouba's qubit takes on one of the four superposition states:
\begin{eqnarray}
&{}&
\alpha\cos\frac{\theta}{2}\,|0\rangle_B + \beta\sin\frac{\theta}{2}e^{i\phi}|1\rangle_B;
\\
&{}&
\alpha\cos\frac{\theta}{2}\,|0\rangle_B - \beta\sin\frac{\theta}{2}e^{i\phi}|1\rangle_B;
\\
&{}&
\beta\cos\frac{\theta}{2}\,|1\rangle_B + \alpha \sin\frac{\theta}{2}e^{i\phi}|0\rangle_B;\\
&{}& 
\beta\cos\frac{\theta}{2}\,|1\rangle_B - \alpha\sin\frac{\theta}{2}e^{i\phi}|0\rangle_B.
\end{eqnarray}
Bouba's qubit is now in a state that resembles the state to be teleported. The four possible states for Bouba's qubit are unitary images of the state to be teleported. After Bouba receives the message from Amy, he guesses which of the four states his qubit is in.  Using this information, Bouba accordingly chooses one of the unitary transformation $\{ \mathbb{I}, \sigma_x,\, i\sigma_y,\, \sigma_z \}$, to perform his part of the channel. Here 
\begin{equation}
\mathbb{I} = \left(
\begin{array}{ccc}
1 & {} & 0\\
0 & {} & 1
\end{array}
\right);\quad
\sigma_x = \left(
\begin{array}{ccc}
0 & {} & 1 \\
1  & {} & 0
\end{array}
\right);\quad
\sigma_y = \left(
\begin{array}{ccc}
0 & {} & -i\\
i & {} & 0
\end{array}
\right);\quad
\sigma_z = \left(
\begin{array}{ccc}
1 & {} & 0 \\
0  & {} & -1
\end{array}
\right).
\end{equation}
$\mathbb{I}$ represents the identity operator,  and $\sigma_x, \sigma_y, \sigma_z$ are the Pauli operators. The correspondence between the measurement outcomes and the unitary operations are 
\begin{equation}
|\phi^+\rangle_{CA} \Rightarrow \mathbb{I};\quad
|\phi^-\rangle_{CA} \Rightarrow \sigma_z;\quad
|\psi^+ \rangle_{CA} \Rightarrow \sigma_x;\quad
|\psi^-\rangle_{CA} \Rightarrow i\sigma_y\,.
\end{equation}
The teleportation is achieved,  and to measure the efficiency of the teleportation protocol,  we compute the fidelity of this teleportation

\subsection{Quantum teleportation fidelity}

\subsubsection{Definition}

The measure of the quality of the teleported state, called fidelity, is similar to the notion of quantum fidelity, which amounts to considering the behavior in time of the overlap of two quantum states: one evolved according to a given dynamics, and the other one evolved by a slight perturbation of it, but starting from the same initial state at time zero. A common definition of the fidelity between two quantum states $\sigma_1$ and $\sigma_2$, expressed as density matrices, is as follows \cite{jozsa}.
\begin{equation}\label{fidel}
F(\sigma_1, \sigma_2) = \left(  
\textrm{Trace}[\;\left(\sqrt{\sigma_1}\sigma_2 \sqrt{\sigma_1}\right)^{\frac{1}{2}}\;]
\right) ^2.
\end{equation}
In quantum teleportation, the fidelity is commonly defined as in the equation (\ref{fidel}). However there are alternative definitions, for instance in \cite{niesen} fidelity is defined as 
\begin{equation}\label{fidel1}
F(\sigma_1, \sigma_2) =   
\textrm{Trace}[\;\left(\sqrt{\sigma_1}\sigma_2 \sqrt{\sigma_1}\right)^{\frac{1}{2}}\;].
\end{equation}
There are other definitions of fidelity as follows.
\begin{itemize}
\item \begin{equation}
F = \langle \psi |\rho | \psi \rangle\,,
\end{equation}
that is the overlap of the input information state  $|\psi \rangle$ with the normalized output teleported state $\rho$.
\item
\begin{equation}
F = \textrm{Tr}[\rho_{\textrm{out}}| \phi\rangle\langle \phi |],
\end{equation}
where $|\phi\rangle$ is the input information state, and $\rho_{\textrm{out}}$, the output information state.
\item The following definition has been used in \cite{prakash1}
\begin{equation}
 F = |\langle T | I \rangle |^2,
\end{equation}
where $|I \rangle$ represents the information state, to be teleported, and $|T\rangle$ represents the teleported copy of the initial information state that the receiver has after application of the unitary transformation. 
\end{itemize}
In this work, we adopt the following definition of fidelity as discussed in \cite{mitali}, which is compatible with the teleportation protocol.
\begin{equation}\label{fed}
F = \sum_{i =1}^4 P_i\,|\,\langle I \,|\,\zeta_i\rangle\,|^2, 
\end{equation}
$|I\rangle$ is the input state and 
$P_i = \textrm{Tr}(\langle \Omega | M_i |\Omega\rangle )$ with $|\Omega \rangle = |I\rangle \otimes |\psi_{channel}\rangle$
$ M_i  = |\psi_i\rangle\langle \psi_i | $ is a measurement operator on a Bell basis.
$|\zeta_i\rangle $ is the teleported state corresponding to $i^{th}$ projective measurement on a Bell basis, where $F$ depends on the parameters of the state to be teleported.
In some references, $|\,\langle I \,|\,\zeta_i\rangle\,|^2$ is considered as fidelity and F in the equation (\ref{fed}) as average fidelity \cite{popescu}.

The quality of the teleportation can also be studied through other measures such as the minimum assured fidelity (MASF), the average teleportation fidelity ($F_{\textrm{ave}}$), and the optimal fidelity $(f)$.  We do not discuss these measures here. Although the many definitions of fidelity, it satisfies the following properties:
i) the fidelity is state-dependent, and ($0\le \textrm{F} \le 1$), ii) if the output state is the same as the input information, then the fidelity of the teleportation is equal to unity, ($ \textrm{F} = 1$), iii) if the output state is imprecise copy of the input information, then the fidelity is smaller than $1,\; (\textrm{F} < 1)$, vi) if the output state is completely orthogonal to the input state, then the fidelity is zero, ($ \textrm{F} = 0$) and the teleportation is not possible, v) the less entangled states reduce the fidelity of teleportation \cite{popescu, horodecki, henderson}.

\subsubsection{Computation of the fidelity}

To measure the efficiency of the teleportation protocol,  we compute the fidelity of this teleportation

\begin{equation}\label{fide}
F (|\psi\rangle_C ) = \sum_{i = 1}^4 P_i\,|\langle\psi_C |\zeta_i \rangle|^2\,,
\end{equation}
where
$P_i = \textrm{Tr}({}_{CAB}\langle \psi |M_i|\psi\rangle_{CAB})$, $M_i = | \phi_i\rangle \langle \phi_i |$  the measurement operator on the quasi-Bell basis 
$|\phi_i \rangle \in \{ |\phi^+ \rangle_{CA},\:  |\phi^-\rangle_{CA},\: |\psi^+\rangle_{CA},\: |\psi^-\rangle_{CA} \},$
and  $|\zeta_i \rangle $ Bouba's normalized and corrected outcome given the measurement result in $i$.

The probabilities of the measurement are explicitly the following:
\begin{eqnarray}\nonumber
P_1 & = & \textrm{Tr}({}_{CAB}\langle \psi | \phi_1\rangle_{}\langle \phi_1 | \psi\rangle_{CAB})  = 
\textrm{Tr}({}_{CAB}\langle \psi | \phi^+_{CA}\rangle_{}\langle \phi^+_{CA} | \psi\rangle_{CAB})
;\\\nonumber
P_2 & = & \textrm{Tr}({}_{CAB}\langle \psi | \phi_2\rangle\langle \phi_2 | \psi\rangle_{CAB})
= \textrm{Tr}({}_{CAB}\langle \psi | \phi_{CA}^-\rangle\langle \phi_{CA}^- | \psi\rangle_{CAB})
;\\\nonumber
P_3 & = & \textrm{Tr}({}_{CAB}\langle \psi | \phi_3\langle \phi_3| \psi\rangle_{CAB}) = 
\textrm{Tr}({}_{CAB}\langle \psi | \psi_{CA}^+\langle \psi_{CA}^+| \psi\rangle_{CAB})
;\\\nonumber
P_4 & = & \textrm{Tr}({}_{CAB}\langle \psi | \phi_4\rangle\langle \phi_4| \psi\rangle_{CAB})
=  \textrm{Tr}({}_{CAB}\langle \psi | \psi_{CA}^-\rangle\langle \psi_{CA}^-| \psi\rangle_{CAB}).
\end{eqnarray}
The probability of Amy measuring $|\phi^+\rangle_{CA}$ or $|\phi^-\rangle_{CA}$ 
is
\begin{equation}\label{pob1}
 P_1 = P_2 = \frac{1}{2}\left( 
\alpha^2\cos^2\frac{\theta}{2} + \beta^2\sin^2\frac{\theta}{2} 
 \right);
\end{equation}
The probability of Amy measuring $|\psi^+\rangle_{CA}$ or 
$|\psi^-\rangle_{CA}$
is
\begin{equation}\label{pob2}
 P_3 = P_4 = \frac{1}{2}\left( 
\alpha^2\sin^2\frac{\theta}{2} + \beta^2\cos^2\frac{\theta}{2} 
 \right).
\end{equation}
The normalized qubits of Bouba after each measurement are given as follows:
\begin{itemize}
\item for the mesurement of $|\phi^+\rangle_{CA}$, Bouba applies the unit operator to $\alpha\cos\frac{\theta}{2}\,|0\rangle_B + \beta\sin\frac{\theta}{2}e^{i\phi}|1\rangle_B$ and normalizes to obtain
\begin{equation}\label{bou1}
|\zeta_1 \rangle = \frac{1}{\sqrt{\alpha^2\cos^2\frac{\theta}{2} + \beta^2\sin^2\frac{\theta}{2}}}\left(
\alpha\cos\frac{\theta}{2}|0\rangle_B  + \beta\sin\frac{\theta}{2}e^{i\phi}|1\rangle_B
\right);
\end{equation}
\item 
for the mesurement of $|\phi^-\rangle_{CA}$, Bouba applies the operator $\sigma_x$ 
on $\alpha\cos\frac{\theta}{2}\,|0\rangle_B - \beta\sin\frac{\theta}{2}e^{i\phi}|1\rangle_B$ and after normalization the outcome is
\begin{equation}\label{bou2}
|\zeta_2 \rangle = \frac{1}{\sqrt{\alpha^2\cos^2\frac{\theta}{2} + \beta^2\sin^2\frac{\theta}{2}}}\left(
 -\beta\sin\frac{\theta}{2}e^{i\phi}|0\rangle_B
 + \alpha\cos\frac{\theta}{2}|1\rangle_B
\right);
\end{equation}
\item 
for the mesurement of $|\psi^+\rangle_{CA}$, Bouba applies the operator $i\sigma_y$ 
on $\beta\cos\frac{\theta}{2}\,|1\rangle_B + \alpha \sin\frac{\theta}{2}e^{i\phi}|0\rangle_B$;
 and after normalization the outcome is
\begin{equation}\label{bou3}
|\zeta_3 \rangle = \frac{1}{\sqrt{\alpha^2\sin^2\frac{\theta}{2} + \beta^2\cos^2\frac{\theta}{2}}}\left(
\beta\cos\frac{\theta}{2}|0\rangle_B
-\alpha\sin\frac{\theta}{2}e^{i\phi}|1\rangle_B
\right);
\end{equation}
\item for the mesurement of $|\psi^-\rangle_{CA}$, Bouba applies the operator $\sigma_z$ 
on $\beta\cos\frac{\theta}{2}\,|1\rangle_B - \alpha\sin\frac{\theta}{2}e^{i\phi}|0\rangle_B.$;
 and after normalization the outcome is
\begin{equation}\label{bou4}
|\zeta_4 \rangle = \frac{1}{\sqrt{\alpha^2\sin^2\frac{\theta}{2} + \beta^2\cos^2\frac{\theta}{2}}}\left(
 -\alpha\sin\frac{\theta}{2}e^{i\phi}|0\rangle_B
 -\beta\cos\frac{\theta}{2}|1\rangle_B
\right).
\end{equation}
\end{itemize}
Using the equations in (\ref{pob1}), (\ref{pob2}), and the equations in (\ref{bou1}),
(\ref{bou2}), (\ref{bou3}), (\ref{bou4}) in the equation (\ref{fide}), the fidelity is given by 
\begin{equation}\label{fede}
F (|\psi\rangle_C ) = \cos^4\frac{\theta}{2} 
+ \sin^4\frac{\theta}{2} + \alpha\beta\sin^2\theta.
\end{equation}
For $\alpha\beta = 1/2$, $F (|\psi\rangle_C ) = 1$, in that case, there is perfect teleportation. This situation implies that $\alpha = 1/\sqrt 2$ and $\beta = 1/\sqrt 2$ and then the state (\ref{sab}) is maximally entangled. Assuming the state (\ref{sab}) non maximally entangled state excludes this situation. 

\section{Conclusion}\label{sec5}

In this paper, we review the role of qubits in quantum entanglement and quantum teleportation. The fidelity of teleportation is also discussed. As concluding remarks, the following challenges are raised:
\begin{enumerate}
\item The separability of quantum states is directly linked to unsolved challenges of mathematics concerning linear algebra and geometry, functional analysis and, in particular, the theory of $C^\star$-algebra. The distillability problem, that is, the question whether the state of a composite quantum system can be transformed to an entangled pure state using local operations, is another problem that is related to challenging open questions of modern mathematics.
These unsolved challenges of mathematics responsible for the quantum separability are addressed in the paper by Pawel Horodecki et al, and a paper by Balachandran et al. \cite{pahoro, bala}.
\item Deterministic perfect teleportation is not possible in the case of entangled non-orthogonal coherent states.
\item The fidelity of teleportation depends on the parameters of the initial state to be teleported. For the protocol, the unitary operators used by the receiver are the Pauli matrices. It may be interesting to find the convenient unitary operators that give perfect teleportation.
\item We are interested in analyzing Hardy's experiment inside the scenario of entanglement and fidelity.
\end{enumerate}

{\bf Appendix}
\begin{equation}
|0 \rangle = 
\left( \begin{array}{cc}
1\\ 0
\end{array}
\right); \quad 
|1 \rangle = 
\left( \begin{array}{cc}
0 \\ 1
\end{array}
\right).
\end{equation}
\begin{equation}
|00\rangle = |0\rangle \otimes |0\rangle = 
\left( \begin{array}{c}
1\\ 0 \\0 \\0 
\end{array}
\right); \quad 
|11 \rangle = |1\rangle \otimes |1\rangle = 
\left( \begin{array}{c}
0\\ 0 \\0 \\1
\end{array}
\right).
\end{equation}
\begin{equation}
|\phi \rangle = \frac{1}{\sqrt 2}\left( |00 \rangle + |11 \rangle\right) = \frac{1}{\sqrt 2}
\left( \begin{array}{c}
1 \\ 0 \\0 \\1
\end{array}
\right).
\end{equation}
\begin{equation}
\langle \phi \vert = \frac{1}{\sqrt 2}\left( \langle 00 \vert + \langle 11 \vert \right) = \frac{1}{\sqrt 2} 
\left( \begin{array}{cccc}
1 & 0  & 0  & 1
\end{array}
\right).
\end{equation}
 \begin{equation}
A  = \frac{1}{\sqrt 2}\left(\begin{array}{cc}
1 & 1\\
1 & -1
\end{array} \right);\: A' = \frac{1}{\sqrt 2}\left(\begin{array}{cc}
-1 & 1\\
1 & 1
\end{array} \right);\: B = \left(\begin{array}{cc}
0 & 1\\
1 & 0
\end{array} \right); \: B' = \left(\begin{array}{cc}
1 & 0\\
0 & -1
\end{array} \right).
\end{equation}
\begin{equation}
A \otimes B = \frac{1}{\sqrt 2}
\left(
\begin{array}{cccc}
0 & 1 & 0 & 1\\
1 & 0 & 1 & 0 \\
0 & 1 & 0 & -1 \\
1 & 0 & -1 & 0 
\end{array}\right); \quad 
A \otimes B' = \frac{1}{\sqrt 2}
\left(
\begin{array}{cccc}
1 & 0 & 1 & 0\\
0 & -1 & 0 & -1 \\
1 & 0 & -1 & 0 \\
0 & -1 & 0 & 1 
\end{array}\right).
\end{equation}
\begin{equation}
A'\otimes B = \frac{1}{\sqrt 2}
\left(
\begin{array}{cccc}
0 & -1 & 0 & 1\\
-1 & 0 & 1 & 0 \\
0 & 1 & 0 & 1 \\
1 & 0 & 1 & 0 
\end{array}\right); \quad 
A' \otimes B' = \frac{1}{\sqrt 2}
\left(
\begin{array}{cccc}
-1 & 0 & 1 & 0\\
0 & 1 & 0 & -1 \\
1 & 0 & 1 & 0 \\
0 & -1 & 0 & -1 
\end{array}\right).
\end{equation}
\begin{equation}
( A \otimes B + A \otimes B' + A' \otimes B - A' \otimes B') 
= \frac{1}{\sqrt 2}\left( 
\begin{array}{cccc}
 2 & 0 & 0 & 2 \\
 0 & -2 & 2 & 0\\
 0 & 2 & -2 & 0 \\
 2 & 0 & 0 & 2
\end{array}
\right).
\end{equation}
A straight calculation shows that 
$\langle \phi |A \otimes B + A \otimes B' + A' \otimes B - A' \otimes B' | \phi \rangle  = 2 \sqrt 2$.

\end{document}